\documentclass[12pt, a4paper]{article}
\usepackage{epsf}
\usepackage{cite}
\usepackage{amsmath,amssymb}
\input{colordvi.tex}
\usepackage[usenames,dvipsnames]{color}
\usepackage{graphicx}
\usepackage{comment}
\begin{document}

\newcommand{\lsim}{\stackrel{<}{_\sim}}
\newcommand{\gsim}{\stackrel{>}{_\sim}}
\newcommand{\rem}[1]{{ {\color{red} [$\spadesuit$ \bf #1 $\spadesuit$]} }}

\renewcommand{\theequation}{\thesection.\arabic{equation}}
\renewcommand{\thefootnote}{\fnsymbol{footnote}}
\setcounter{footnote}{0}

%%%%%%%%%%%%%%%%%%%%%%%%%%%%%%%%%%%%%%%%%%%%%%%%%%

%%%%%%%%%%%%%%%%%%%%%%%%%%%%%%%%%%%%%%%%%%%%%%%%%%
\def\thefootnote{\fnsymbol{footnote}}
\def\a{\alpha}
\def\b{\beta}
\def\c{\varepsilon}
\def\d{\delta}
\def\e{\epsilon}
\def\f{\phi}
\def\g{\gamma}
\def\h{\theta}
\def\k{\kappa}
\def\l{\lambda}
\def\m{\mu}
\def\n{\nu}
\def\p{\psi}
\def\q{\partial}
\def\r{\rho}
\def\s{\sigma}
\def\t{\tau}
\def\u{\upsilon}
\def\v{\varphi}
\def\w{\omega}
\def\x{\xi}
\def\y{\eta}
\def\z{\zeta}
\def\D{\Delta}
\def\G{\Gamma}
\def\H{\Theta}
\def\L{\Lambda}
\def\F{\Phi}
\def\P{\Psi}
\def\S{\Sigma}
\def\me{\mathrm e}

\def\o{\over}
\def\beq{\begin{eqnarray}}
\def\eeq{\end{eqnarray}}
\newcommand{\vev}[1]{ \left\langle {#1} \right\rangle }
\newcommand{\bra}[1]{ \langle {#1} | }
\newcommand{\ket}[1]{ | {#1} \rangle }
\newcommand{\bs}[1]{ {\boldsymbol {#1}} }
\newcommand{\mc}[1]{ {\mathcal {#1}} }
\newcommand{\mb}[1]{ {\mathbb {#1}} }
\newcommand{\EV}{ {\rm eV} }
\newcommand{\KEV}{ {\rm keV} }
\newcommand{\MEV}{ {\rm MeV} }
\newcommand{\GEV}{ {\rm GeV} }
\newcommand{\TEV}{ {\rm TeV} }
\def\diag{\mathop{\rm diag}\nolimits}
\def\Spin{\mathop{\rm Spin}}
\def\SO{\mathop{\rm SO}}
\def\O{\mathop{\rm O}}
\def\SU{\mathop{\rm SU}}
\def\U{\mathop{\rm U}}
\def\Sp{\mathop{\rm Sp}}
\def\SL{\mathop{\rm SL}}
\def\tr{\mathop{\rm tr}}
\def\sp{\;\;}

\def\IJMP{Int.~J.~Mod.~Phys. }
\def\MPL{Mod.~Phys.~Lett. }
\def\NP{Nucl.~Phys. }
\def\PL{Phys.~Lett. }
\def\PR{Phys.~Rev. }
\def\PRL{Phys.~Rev.~Lett. }
\def\PTP{Prog.~Theor.~Phys. }
\def\ZP{Z.~Phys. }
%%%%%%%%%%%%%%%%%%%%%%%%%%%%%%%%%%%%%%%%%%%%%%%%%%

%%%%%%%%%%%%%%%%%%%%%%%%%%%%%%%%%%%%%%%%%%%%%%%%%%
\begin{titlepage}

\begin{center}

\hfill UT-15-31\\

\vskip .75in

{\Large \bf 
Curvature Perturbation and Domain Wall Formation\\[.5em]
with Pseudo Scaling Scalar Dynamics
}

\vskip .75in

{\large Yohei Ema$^a$, Kazunori Nakayama$^{a,b}$ and Masahiro Takimoto$^a$}

\vskip 0.25in

\begin{tabular}{ll}
$^{a}$&\!\! {\em Department of Physics, Faculty of Science, }\\
& {\em The University of Tokyo,  Bunkyo-ku, Tokyo 133-0033, Japan}\\[.3em]
$^{b}$ &\!\! {\em Kavli IPMU (WPI), UTIAS,}\\
&{\em The University of Tokyo,  Kashiwa, Chiba 277-8583, Japan}
\end{tabular}

\end{center}
\vskip .5in

\begin{abstract}

Cosmological dynamics of scalar field with a monomial potential $\phi^n$
with a general background equation of state is revisited.
It is known that if $n$ is smaller than a critical value, the scalar field exhibits a coherent oscillation
and if $n$ is larger it obeys a scaling solution without oscillation.
We study in detail the case where $n$ is equal to the critical value, and find a
peculiar scalar dynamics which is neither oscillating nor scaling solution,
and we call it a pseudo scaling solution.
We also discuss cosmological implications of a pseudo scaling scalar dynamics,
such as the curvature perturbation and the domain wall problem.

\end{abstract}

\end{titlepage}

\renewcommand{\thepage}{\arabic{page}}
\setcounter{page}{1}
\renewcommand{\thefootnote}{\#\arabic{footnote}}
\setcounter{footnote}{0}
%%%%%%%%%%%%%%%%%%%%%%%%%%%%%%%%%%%%%%%%%%%%%%%%%%

%%%%%%%%%%%%%%%%%%%%%%%%%%%%%%%%%%%%%%%%%%%%%%%%%%
\section{Introduction}
\setcounter{equation}{0}
%%%%%%%%%%%%%%%%%%%%%%%%%%%%%%%%%%%%%%%%%%%%%%%%%%

In the early universe, a scalar field often oscillates coherently after inflation~\cite{Guth:1980zm}.
Coherent oscillation of a scalar field may have various cosmological effects such as 
generation of curvature perturbation, dark matter, baryon asymmetry and unwanted relics.

In a simple model of a real scalar field $\phi$ with a canonical kinetic term and a scalar potential of $V \propto \phi^n$,
it is known that the scalar field stops at the initial position $\phi_i$ until the Hubble parameter becomes equal to
the effective mass of the scalar.
After that, $\phi$ begins a coherent oscillation for $n<n_c$ with $n_c$ being the critical exponent, which depends
on the background equation of state (Eq.~(\ref{cond1})).
In this case, the mass scale of $\phi$ becomes larger and larger compared with the Hubble scale
and such a coherent scalar dissipates its energy due to the Hubble expansion.
The amplitude of the coherent oscillation $\Phi$ decreases as~\cite{Turner:1983he}
\begin{equation}
	\Phi \propto a^{-6/(n+2)},   \label{scale1}
\end{equation}
with $a(t)$ being the scale factor of the universe.
Such a scalar field can act as a curvaton~\cite{Mollerach:1989hu}, 
which generates the primordial density perturbation of the universe~\cite{Dimopoulos:2003ss}.
In particular, curvaton models with $n>2$ were studied in some literature~\cite{Enqvist:2009zf,Byrnes:2011gh,Mukaida:2014wma}.

On the other hand, for $n>n_c$, the scalar field never oscillates and it just follows a so-called scaling solution~\cite{Liddle:1998xm}\footnote{
	Note on terminology: in this paper we just call Eq.~(\ref{scale2}) scaling solution.
	We do not regard an oscillating solution as a scaling solution even if its amplitude satisfies a simple power law like (\ref{scale1}).
}
\begin{equation}
	\phi \propto t^{-\frac{2}{n-2}}.   \label{scale2}
\end{equation}
This is an attractor and it was pointed out that such a scalar field cannot generate curvature perturbation
since the initial fluctuation of $\phi_i$ is erased during the scaling regime~\cite{Dimopoulos:2003ss}.

In this paper, we revisit the scalar dynamics for $n=n_c$.
In Ref.~\cite{Dimopoulos:2003ss} it is stated that in this critical case, 
the scalar field is not attracted to a scaling solution but it is something like between a scaling and oscillation.
Nevertheless, they say that for $n=n_c$ both the oscillating solution (\ref{scale1}) and scaling solution (\ref{scale2}) show the same behavior
with the cosmic expansion and hence the curvature perturbation also vanishes.
However, we will see that the actual dynamics of the scalar field for $n=n_c$
is far from both the oscillating solution (\ref{scale1}) and scaling solution (\ref{scale2}).
Even averaged features of the scalar dynamics cannot be described by these solutions.
Instead, a peculiar feature of the scalar dynamics shows up, which we call a ``pseudo scaling solution''.
We will show it both in a numerical and analytical way and see that the curvature perturbation does 
not vanish in this case.

The pseudo scaling solution can have a variety of physical implications. As we noted above, 
the curvature perturbation can be generated by a curvaton even if it follows the pseudo scaling solution,
and hence it may offer the seed of the large scale structure of our universe. Another example is a symmetry breaking
field such as the Peccei-Quinn field~\cite{Peccei:1977hh}. In the supersymmetric (SUSY) axion model, it is possible that the 
Peccei-Quinn field follows the pseudo scaling solution in the early universe. As we will see later, 
we can naturally solve the isocurvature perturbation as well as the domain wall problem in such a case.
The pseudo scaling behavior also has some implications on the baryon number asymmetry produced 
by the Affleck-Dine mechanism~\cite{Affleck:1984fy}.
Therefore, it is worth studying detailed features of the pseudo scaling solution.

The organization of this paper is as follows. In Sec.~\ref{sec:mono_pot}, we discuss the dynamics of
the scalar field with a monomial potential in a general background equation of state. 
Here the ``pseudo scaling" scalar dynamics is extensively
studied. Also, some comments on the Hubble mass term are given there. 
Cosmological implications of the pseudo scaling solution are explained in Sec.~\ref{sec:cos}. 
We discuss the curvature perturbation and the domain wall problem there. The last section
is devoted to conclusions and discussion.

%%%%%%%%%%%%%%%%%%%%%%%%%%%%%%%%%%%%%%%%%%%%%%%%%%
\section{Scalar dynamics with monomial potential}
\setcounter{equation}{0}
\label{sec:mono_pot}
%%%%%%%%%%%%%%%%%%%%%%%%%%%%%%%%%%%%%%%%%%%%%%%%%%

%%%%%%%%%%%%%%%%%%%%%%%%%%%%%%%%%%%%%%%%%%%%%%%%%%
\subsection{Scaling and oscillating solution for $n\neq n_c$}
%%%%%%%%%%%%%%%%%%%%%%%%%%%%%%%%%%%%%%%%%%%%%%%%%%

Let us consider the Lagrangian
\begin{align}
	\mathcal L = -\frac{1}{2}(\partial \phi)^2 - V(\phi),
\end{align}
where the scalar potential is given by the monomial one
\begin{equation}
	V(\phi) =  \frac{\lambda}{n}\phi^n,
\end{equation}
where $n>2$ is an integer.\footnote{
For an odd $n$, $\phi$ should be interpreted as $|\phi|$.}
The equation of motion of $\phi$ is given by
\begin{equation}
	\ddot\phi + 3H\dot\phi + \lambda \phi^{n-1} = 0,
\end{equation}
with $H$ being the Hubble parameter.
We take $H=p/t$ and assume $3p-1>0$. In the matter (radiation)-dominated (MD (RD)) universe, $p=2/3\,(1/2)$. 
A more general equation of state may be obtained in the inflaton oscillation dominated universe (see App.~\ref{sec:app}).
Let us define
\begin{align}
	\varphi \equiv a^{\frac{2}{p(n-2)}}\phi,\\
	s\equiv \ln(t/t_i),
\end{align}
with $a(t)$ being the scale factor of the universe and $t_i$ the initial time. We have taken $a(t_i)=1$.
Then the equation of motion becomes~\cite{Dine:1995kz,Harigaya:2015hha}
\begin{align}
	\varphi'' + \left(3p-\frac{n+2}{n-2}\right)\varphi' +\mu^2\varphi + \frac{\lambda p^2}{H_i^2} \varphi^{n-1} = 0,  \label{eqofm1}
\end{align}
where the prime denotes derivative with respect to $s$ and
\begin{equation}
	\mu^2 \equiv \frac{2(6p-3pn+n)}{(n-2)^2}.
\end{equation}
The equation (\ref{eqofm1}) represents the motion of the rescaled field $\varphi$ in the effective potential
\begin{align}
	V_{\rm eff}(\varphi)=\frac{1}{2}\mu^2\varphi^2 + \frac{\lambda p^2}{nH_i^2} \varphi^{n}. \label{Veff}
\end{align}
In these variables, $\varphi$ has a minimum at $\varphi_{\rm min}$ where
\begin{align}
	\varphi_{\rm min} = \begin{cases}
		0 &{\rm if}~~~\mu^2 > 0 \\
		(-\mu^2H_i^2/\lambda p^2)^{1/(n-2)} &{\rm if}~~~\mu^2 < 0.
	\end{cases}
\end{align}

Thus if
\begin{equation}
	n > n_c\equiv \frac{6p+2}{3p-1}=\begin{cases}
		4 &{\rm for}~~~p=1\\
		6 &{\rm for}~~~p=2/3 \\
		10 &{\rm for}~~~p=1/2
	\end{cases}
	~~~\leftrightarrow~~~ p = \frac{n_c+2}{3(n_c-2)}
	, \label{cond1}
\end{equation}
the sign of the friction term is positive, and hence the oscillation of $\varphi$ damps and
it relaxes to the minimum of the potential $\varphi_{\rm min}$.

Let us suppose the condition (\ref{cond1}) is satisfied so that there is a nontrivial minimum of the effective potential
and $\varphi$ relaxes there.\footnote{
	Note that $\mu^2 < 0 \Leftrightarrow n_{c} + 2 < 2n$, and $n_{c} > 2$ for $p > 1/3$.
	Thus, $\mu^2 < 0$ is ensured and there is a nontrivial minimum of the 
	potential if $n \geq n_{c}$.
	It does not always hold if there is a Hubble mass term in the potential (see Eq.~\eqref{eq:mass_h}).
}
This represents the scaling solution: in terms of the original field $\phi$,
\begin{equation}
	\phi(t) \propto a^{-\frac{2}{p(n-2)}} \propto t^{-\frac{2}{n-2}}.  \label{scaling}
\end{equation}
In this case, independently of the initial condition, $\phi$ finally follows the scaling solution without oscillation.
Note also that the scaling solution implies that the effective mass scale of the scalar decreases in proportion to the Hubble parameter:
$m_{\rm eff}^2 = \lambda \phi^{n-2} \propto H^2$.

On the other hand, if the inequality (\ref{cond1}) is inverted, the friction term of (\ref{eqofm1}) has a negative sign,
and hence $\varphi$ never relaxes to its potential minimum.
This corresponds to an oscillating solution.
This behavior is also understood as follows.
Assuming an oscillating solution for $\phi$ with potential $\sim \phi^n$, we obtain 
\begin{equation}
	\Phi(t) \propto a^{-6/(n+2)} \propto t^{-6p/(n+2)},
\end{equation}
by making use of the Virial theorem
with $\Phi$ being the oscillation amplitude of $\phi$.
Thus the effective mass of $\phi$ scales as $m_{\rm eff} \propto \Phi^{(n-2)/2} \propto t^{-3p(n-2)/(n+2)}$.
If this can continue to exceed the Hubble parameter $H \sim t^{-1}$, the oscillation survives.
This condition is written as
\begin{equation}
	\frac{3p(n-2)}{(n+2)} < 1 ~~\leftrightarrow~~ n <  \frac{6p+2}{3p-1}.
\end{equation}
This is just an inversion of the condition (\ref{cond1}).

%%%%%%%%%%%%%%%%%%%%%%%%%%%%%%%%%%%%%%%%%%%%%%%%%%
\subsection{Pseudo scaling solution for $n=n_c$}
\label{sec:PS}
%%%%%%%%%%%%%%%%%%%%%%%%%%%%%%%%%%%%%%%%%%%%%%%%%%

We have seen that a scalar field approaches to a scaling solution for $n> n_c$ and oscillating solution for $n< n_c$
independently of the initial condition.
However, the case of $n=n_c$ is more nontrivial.
In this case, the friction term of (\ref{eqofm1}) vanishes:
\begin{align}
	\varphi'' -\frac{4}{(n-2)^2}\varphi + \frac{(n+2)^2}{9(n-2)^2}\frac{\lambda}{H_i^2} \varphi^{n-1} = 0.  \label{eqofm2}
\end{align}
In terms of $\varphi$, it does not universally approach to some fixed solutions and continues to oscillate without damping.
The resulting dynamics significantly depends on the initial condition of $\phi$.
If the initial condition is chosen so that $\varphi$ oscillates around $\varphi_{\rm min}$ without crossing $\varphi=0$,
$\phi$ also does not cross $\phi=0$. 
The resulting dynamics of $\phi$ looks like neither scaling nor oscillating solution;
we call this  a ``pseudo scaling solution''.
Hereafter we consider the case of $n=n_c$.

Before solving the equation of motion, we must specify the initial condition.
The natural initial condition is set just after inflation as $\phi(t_i)=\phi_i$ and $\dot \phi(t_i)=\dot\phi_i \simeq -\partial_{\phi_i}V(\phi_i)/3H_i$,
where $H_{i}$ and $t_{i}$ are the Hubble parameter and the cosmic time at the end of the inflation, respectively.
We assume $\phi_{i} > 0$ without loss of generality.
Here $\phi_i$ should satisfy
\begin{align}
	m_{\rm eff}^2(\phi_i) \equiv \lambda \phi_i^{n-2} \lesssim H_i^2,
\end{align}
since otherwise $\phi$ rolls down the potential during inflation.
This condition roughly means $\varphi_i \lesssim \varphi_{\rm min}$.
Therefore, in terms of $\varphi$, it is initially located close to the origin and rolls down the effective potential (\ref{Veff}) toward the minimum.
In particular, we consider the case 
\begin{equation}
	\varphi_i \ll \varphi_{\rm min}.
\end{equation}
By noting that the initial velocity of $\varphi$ is given by
\begin{equation}
	\varphi_i' = t_i\left( \dot \phi_i + \frac{6}{n+2}H_i\phi_i \right),
\end{equation}
we obtain the conserved energy density of $\varphi$:
\begin{align}
	\rho_\varphi = \frac{1}{2}\varphi_i'^2 + V_{\rm eff}(\varphi_i) &\simeq 
	\frac{n+2}{3(n-2)^2 H_i}\left(2\phi_i\dot\phi_i + \frac{n+2}{3n}\frac{\lambda\phi_i^n}{H_i} \right) \nonumber\\
	&\simeq -\frac{n+2}{9n(n-2)}\frac{\lambda \phi_i^n}{H_i^2}.
	\label{eq:energy}
\end{align}
Here we have used the approximation $|\dot\phi_i|\ll |H_i\phi_i|$ which is valid as long as $\varphi_i \ll \varphi_{\rm min}$.
Since the energy density is negative, $\varphi$ does not overshoot the origin $\varphi=0$,
which means that $\phi$ does not oscillate around $\phi=0$.
The rescaled field $\varphi$ just oscillates around $\varphi_{\rm min}$ without damping.\footnote{
	This observation crucially depends on our assumption on the initial condition $3H_i\dot\phi_i\simeq -V'$.
	Although it holds during inflation, the relation is modified after inflation depending on the effective equation of state of 
	the inflaton oscillation dominated universe as $H\dot\phi \simeq -pV'/(3p+1)$ for $|V''/H^2|\ll 1$.
	The approximation $3H_i\dot\phi_i\simeq -V'$ as an initial condition is valid for inflation models 
	in which the inflaton mass scale is much larger than the Hubble scale, such as new inflation. 
	We have numerically checked that $\rho_\varphi$ is indeed negative for such a case.
	On the other hand, for chaotic inflation, our approximation is not justified and $\rho_\varphi$ may be positive.
	However, the dynamics is almost identical to the case of negative $\rho_\varphi$ 
	after $\phi$ is replaced with $|\phi|$ even if $\phi$ once overshoots $\phi=0$ as long as
	$\dot{\phi_{i}} \sim \mathcal{O}(V^{\prime}/H_{i})$ is satisfied.
}

%%%%%%%%%%%%%%%%
\begin{figure}[t]
\begin{center}
\includegraphics[scale=1.0]{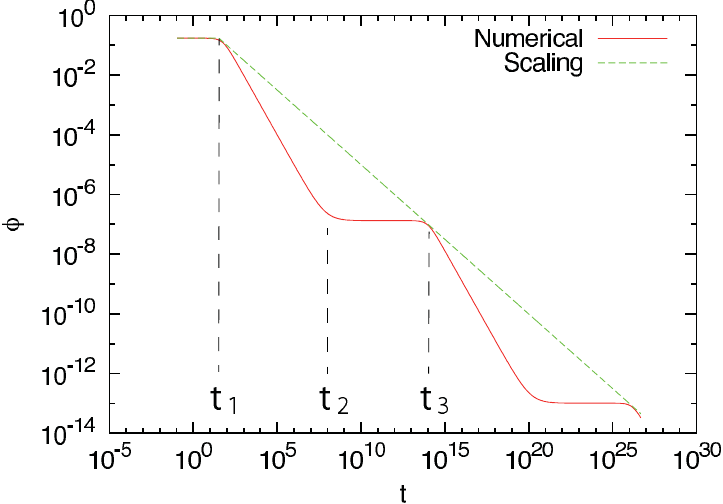}
\includegraphics[scale=1.0]{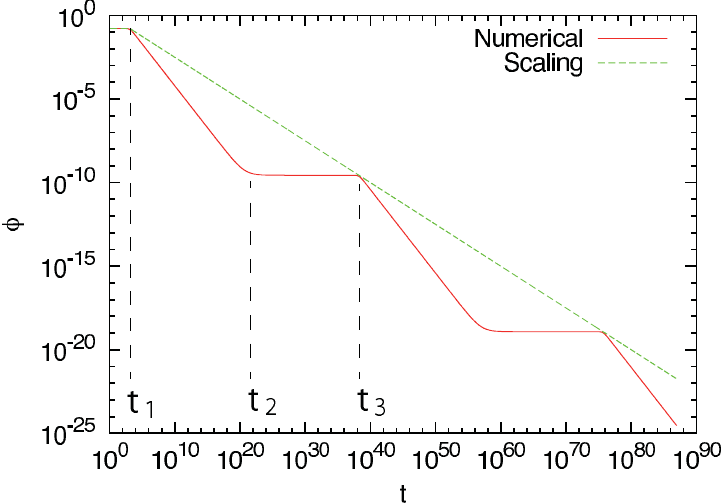}
\end{center}
\caption {
	Time evolution of $\phi(t)$ for $n=n_c$ for the potential $\phi^n$
	for $p=2/3\,(n=6)$ in the left panel and $p=1/2\,(n=10)$ in the right panel.
	We have normalized $\phi$ by the Planck mass $M_{P}$. The time is normalized by
	$\left(\lambda M_{P}^{n-2}\right)^{-1/2}$, where $\left(\lambda M_{P}^{n-2}\right)^{1/2}$ is  a natural mass scale of this system.
	The red solid line represents a numerical solution, which corresponds to a ``pseudo scaling'' solution,
	while the green dashed line shows a hypothetical scaling solution for representative purpose.
}
\label{fig:PS}
\end{figure}
%%%%%%%%%%%%%%%%

%%%%%%%%%%%%%%%%
\begin{figure}[t]
\begin{center}
\includegraphics[scale = 1.3]{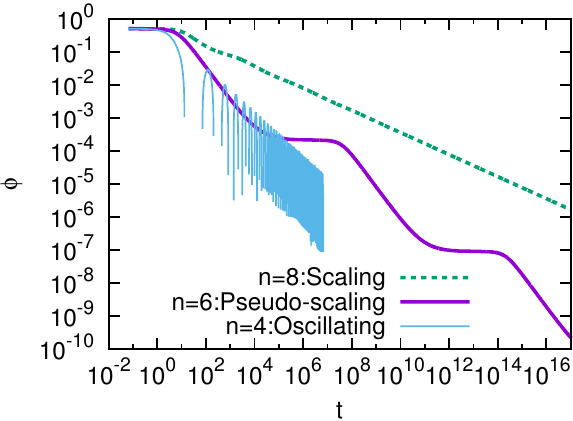}
\end{center}
\caption {
	Time evolution of $\phi(t)$ for $n = 8, 6$ and 4 for $p=2/3\,(n_{c}=6)$.
	The green dashed line is the solution for $n=8$ (scaling). The purple line
	corresponds to the solution for $n=6$ (pseudo-scaling), and the cyan line is
	the solution for $n=4$ (oscillating).
}
\label{fig:PS2}
\end{figure}
%%%%%%%%%%%%%%%%

The dynamics looks so simple in terms of the rescaled field $\varphi$.
However, this ``pseudo scaling'' solution is more nontrivial in terms of the original field $\phi$.
Let us see Fig.\,\ref{fig:PS} where we have numerically solved the equation of motion of $\phi$ with initial conditions 
$H_i=1, \phi_i=0.2\varphi_{\rm min}, \dot\phi_i=-V'(\phi_i)/3H_i$ for $p=2/3\,(n=6)$ in the left panel and $p=1/2\,(n=10)$ in the right panel.
We have normalized $\phi$ by the Planck mass $M_{P}$. The time is normalized by
$\left(\lambda M_{P}^{n-2}\right)^{-1/2}$, where $\left(\lambda M_{P}^{n-2}\right)^{1/2}$ is  a natural mass scale of this system.
The red solid line corresponds to a numerical solution, which corresponds to a pseudo scaling solution,
while the green dashed line shows a scaling solution $\phi \propto t^{-2/(n-2)}$ for representative purpose.
For comparison, we also show time evolution of $\phi(t)$ for cases with $n=8$ (scaling), $n=6$ (pseudo scaling)
and $n=4$ (oscillating) for $p=2/3\, (n_{c} = 6)$ in Fig.~\ref{fig:PS2}. It is clear from this figure that
the pseudo scaling solution resembles neither the scaling solution nor the oscillating solution even qualitatively.
To understand this behavior, let us solve the equation (\ref{eqofm2}) more carefully.

Our goal is to understand main features of the pseudo scaling solution and estimate
how the periodicity depends on the initial condition $\phi_{i}$.\footnote{
We can formally solve Eq.~\eqref{eqofm2}
(see App.~\ref{sec:app2}), but the following discussion is enough for our purpose.
}
Here it is important to note that the dynamics of $\varphi$ is dominated by the region $\varphi \ll \varphi_{\rm min}$
since $|\rho_{\varphi}| \ll |V(\varphi_{\rm min})|$ in the case of our interest.\footnote{
This also holds for other initial conditions as long as the relation $\dot{\phi}_{i} \sim \mathcal{O}\left(V^{\prime}/H_{i}\right)$
is satisfied.}
It means that we can almost neglect
the second term of Eq.\eqref{Veff} in order to understand main features of the dynamics.
In the region $\varphi \ll \varphi_{\rm min}$, the equation of motion is approximated as
\begin{equation}
	\varphi'' - \frac{4}{(n-2)^2}\varphi = 0.
\end{equation}
This has a general solution consisting of the growing and decaying mode,
\begin{equation}
	\varphi(s) = A \exp\left(\frac{2s}{n-2}\right) + B \exp\left(\frac{-2s}{n-2}\right),  \label{sol}
\end{equation}
where $A$ and $B$ are determined by the initial conditions.
In terms of $\phi$, this solution corresponds to
\begin{equation}
	\phi(t) = A + B\left( \frac{t_i}{t} \right)^{4/(n-2)}.   \label{solphi}
\end{equation}
Let us divide the time into the first phase $t_i < t < t_1$, 
the second phase $t_1 < t < t_2$ and the third phase $t_2 < t < t_3$.
Here, we define $t_1$ as the time at which the motion of $\phi$ starts. The time when the motion of $\phi$ stops
is denoted as $t_2$. Some time after $t=t_2$ the field $\phi$ again starts to move, and we define that time as $t_{3}$ (see Fig.\,\ref{fig:PS}).
The $\varphi^{n}$ term in the effective potential becomes important only for some short periods at around $t = t_1, t_2$ and $t_3$.

\paragraph{First phase}
In the time interval $t_i<t<t_1$, $A$ and $B$ are determined by the initial conditions:
\begin{align}
	A=\frac{1}{2}\left( \varphi_i + \frac{n-2}{2}\varphi_i' \right),~~~B=\frac{1}{2}\left( \varphi_i - \frac{n-2}{2}\varphi_i' \right).
\end{align}
This solution breaks down at $s\simeq s_1\,(t\simeq t_1)$ where $\varphi$ becomes close to $\varphi_{\rm min}$.
After that $\varphi$ passes through $\varphi_{\rm min}$ and goes back to $\sim \varphi_{\rm min}$ within order one unit time in terms of $s$.
We can estimate $s_1$ and $t_1$ as
\begin{equation}
	s_1 \simeq \frac{n-2}{2}\ln\left(\frac{\varphi_{\rm min}}{\varphi_i}\right)~~~\leftrightarrow~~~
	t_1 \simeq t_i\left(\frac{\varphi_{\rm min}}{\varphi_i}\right)^{(n-2)/2} \simeq \left(\frac{1}{\lambda\varphi_i^{n-2}}\right)^{1/2}=\frac{1}{m_{\rm eff}(\phi_i)}.  \label{t1}
\end{equation}
Thus the time $t_1$ is roughly the epoch at which the Hubble parameter becomes equal to the effective mass of $\phi$.
It is obvious that $\phi$ remains almost constant during this period.

\paragraph{Second phase}
Next, let us consider the interval $t_1 < t < t_2$.
Just what we have to do is to determine the constants $A$ and $B$ in (\ref{sol}) with appropriate initial conditions.
We set the initial condition at $\varphi=\varphi_j$ (and $s = s_j$), with $\varphi_j$ taken somewhere between $\varphi_i$ and $\varphi_{\rm min}$,
say, $\varphi_j \sim 0.1\varphi_{\rm min}$.
The precise position of $\varphi_j$ is not important since the dynamics is dominated by the region $\varphi \ll \varphi_{\rm min}$ as 
we noted before.
Since $\rho_\varphi$ is conserved during the whole process, we can express $\varphi_j'$ as
\begin{equation}
	\varphi_j' =-\left( 2\rho_\varphi + \frac{4}{(n-2)^2}\varphi_j^2 \right)^{1/2} \simeq \frac{-2}{n-2}\varphi_j(1-\epsilon),
\end{equation}
where
\begin{equation}
	\epsilon \equiv \frac{n^2-4}{36n}\frac{\lambda\varphi_i^n}{H_i^2 \varphi_j^2}
	=\frac{n-2}{n(n+2)}\frac{\varphi_i^n}{\varphi_j^2\varphi_{\rm min}^{n-2}}   \ll 1.
\end{equation}
The sign of $\varphi^{\prime}$ is flipped because $\varphi$ changes the direction of motion due to the higher order term 
in $V_{\rm eff}(\varphi)$ at $t \simeq t_1$.
Then, defining $A_j$ and $B_j$ as the constants $A$ and $B$ in Eq.~\eqref{sol} at $s = s_{j}$, we obtain
\begin{align}
	A_j\simeq \frac{1}{2}\varphi_j \epsilon,~~~B_j\simeq \varphi_j,
\end{align}
hence $A_j \ll B_j$.
Then the end of the second period, $s=s_2$, is the time at which the decaying mode $\sim B_j e^{-2(s_2-s_j)/(n-2)}$ becomes equal to
the growing mode $\sim A_je^{2(s_2-s_j)/(n-2)}$.
Thus we obtain $s_2\,(t_2)$ as
\begin{equation}
	s_2-s_j \simeq \frac{n-2}{4}\ln\frac{2}{\epsilon},~~~\leftrightarrow~~~t_2 \simeq c\,t_1 \left( \frac{\varphi_{\rm min}}{\varphi_i} \right)^{n(n-2)/4},
	\label{eq:t2}
\end{equation}
where in the last equality we have approximated $t_j\sim t_1$ and $\varphi_j \sim \varphi_{\rm min}$.
Here we have included an $\mathcal{O}(1\mathchar`-10)$ constant $c$ which comes from the uncertainty of $\varphi_j$ as well as effects
of the $\varphi^{n}$ term. We have numerically checked that the $\varphi_i$ dependence of $t_2$ is precisely given by Eq.~\eqref{eq:t2}
for $\varphi_i \ll \varphi_{\rm min}$, or $m_{\rm eff}(\phi_i) \ll H_i$.
It should be noticed that $t_2$ is orders of magnitude longer than the inverse mass scale of the scalar field.
From (\ref{solphi}), we can see that $\phi$ scales as $\propto t^{-4/(n-2)}$ during this time interval, so we have
\begin{align}
	\phi(t) \simeq \varphi_i \left( \frac{t_1}{t} \right)^{4/(n-2)}~~~{\rm for}~~~t_1<t<t_2 ,
\end{align}
and eventually at $t=t_2$ it becomes 
\begin{equation}
	\phi(t_2) \sim \varphi_{i} \left( \frac{\varphi_i}{\varphi_{\rm min}} \right)^{n}.
	\label{phit2}
\end{equation}
Note that during this regime, $\phi$ satisfies $\ddot\phi+3H\dot\phi\simeq 0$, so the potential term becomes irrelevant
and the effective mass scale $m_{\rm eff}(\phi)$ becomes smaller and smaller than the Hubble parameter.
The energy density of $\phi$ is dominated by the kinetic term and  scales as $\rho_\phi \propto a^{-6}$ during this regime.

\paragraph{Third phase}
Lastly, let us consider the third period $t_2 < t < t_3$.
This is just a continuation of the second period, where the growing mode dominates over the decaying mode.
Thus the solution is
\begin{equation}
	\varphi(s) \simeq \frac{1}{2}\varphi_j \epsilon \exp\left(\frac{2}{n-2}(s-s_j)\right).
\end{equation}
This is valid until it becomes close to $\varphi_{\rm min}$.
Thus the end of the third period, $s_3\,(t_3)$ is estimated as\footnote{
The constant $c$ here is exactly the same as that in Eq.~\eqref{eq:t2} since the dynamics of $\varphi$ is periodic.
}
\begin{equation}
	s_3-s_2 \simeq \frac{n-2}{4}\ln\frac{2}{\epsilon},~~~\leftrightarrow~~~t_3 \simeq c\,t_2 \left( \frac{\varphi_{\rm min}}{\varphi_i} \right)^{n(n-2)/4}
	\simeq \frac{1}{m_{\rm eff}(\phi(t_2))}.
\end{equation}
Therefore it is seen that the relative time interval $t_3/t_2$ is equal to $t_2/t_1$.
During this time interval, $\phi(t) = \phi(t_2)$ remains almost constant as is clear from (\ref{solphi}).

\paragraph{}
After $t=t_3$, we have exactly the same dynamics as the second phase and the following third phase
since there is no damping in the equation of motion of $\varphi$.
This one ``pseudo-oscillation'' consisting of one set of the second and third period
takes much longer time than the inverse mass scale of the scalar field and even than the Hubble time scale.
In terms of $\phi$, the field value reduces by an amount of $\epsilon \simeq(\varphi_i/\varphi_{\rm min})^{n}$ during this one pseudo oscillation.
The scaling of $\phi$ consists of $\phi= {\rm const}$ regime and $\phi\propto t^{-4/(n-2)}$ regime,
both are completely different from the scaling solution $\propto t^{-2/(n-2)}$.
The scaling solution is obtained only after averaging the pseudo oscillation, but the period of the pseudo oscillation is
much longer than the Hubble time scale, hence using the scaling solution would lead to qualitatively wrong conclusions on the physical implications
such as the curvature perturbation.

Finally, we make some comments on qualitative differences between the scaling solution for $n > n_c$
and  the pseudo scaling solution for $n = n_c$. In the case of the scaling solution, the kinetic term, 
the Hubble friction term and the potential term in the equation of motion are all comparable.
Therefore, if the potential is not monomial, the motion of $\phi$ soon deviates from 
the scaling solution once other terms than the original one dominate the potential.
On the other hand, in the case of the pseudo scaling solution,
the kinetic term and the Hubble friction term dominate over the potential term.
This can be seen by the fact that $\phi = {\rm const}$ and $\phi \propto t^{-4/(n-2)}$ are the solutions of the equation
\begin{align}
\ddot{\phi} + 3H\dot{\phi} = 0,
\end{align}
for $p = (n+2)/3(n-2)$.\footnote{
In terms of $\varphi$, it means that the quadratic term dominates the effective potential.
}
Thus, even after the potential is dominated by other terms than the one with $n = n_c$, 
the motion of $\phi$ still follows the pseudo scaling solution for some period until
the potential term becomes comparable to the kinetic and friction terms.\footnote{
If $p$ changes, e.g. due to the inflaton decay, the pseudo scaling behavior terminates soon.
}We will see this in more detail in Sec.~\ref{sec:cos}.

%%%%%%%%%%%%%%%%%%%%%%%%%%%%%%%%%%%%%%%%%%%%%%%%%%
\subsection{Comment on the case of Hubble mass}  \label{sec:com}
%%%%%%%%%%%%%%%%%%%%%%%%%%%%%%%%%%%%%%%%%%%%%%%%%%

Here we briefly comment on the case where the Hubble mass term exists in the potential:
\begin{equation}
	V(\phi) =  \frac{c}{2}H^2\phi^2 + \frac{\lambda}{n}\phi^n.
\end{equation}
The equation of motion of $\phi$ is given by
\begin{equation}
	\ddot\phi + 3H\dot\phi + cH^2\phi + \lambda \phi^{n-1} = 0.
\end{equation}
The equation of motion of the rescaled field $\varphi$ is the same as (\ref{eqofm1}) except that $\mu^2$ is now replaced with
\begin{equation}
	\mu^2 \equiv \frac{2(6p-3pn+n)}{(n-2)^2}+cp^2.
	\label{eq:mass_h}
\end{equation}
In particular, for $n=n_c$, the equation of motion becomes
\begin{align}
	\varphi'' +\left( \frac{-36+c(n+2)^2}{9(n-2)^2} \right)\varphi + \frac{(n+2)^2}{9(n-2)^2}\frac{\lambda}{H_i^2} \varphi^{n-1} = 0.
\end{align}
As long as $|c|\ll 1$, the existence of the Hubble mass term does not qualitatively affect the pseudo scaling dynamics 
studied in the previous subsection.
If $|c| \gtrsim 1$, the crucial difference from the case of $|c|\ll 1$ is that the initial position of $\phi$ cannot be taken freely.
In particular, for positive $c \gtrsim 1$, we have $\phi_i=0$ and there is no dynamics.

Let us consider the case of negative Hubble mass: $c \lesssim -1$.
In this case, $\phi$ initially sits at the minimum of the potential $\phi=\phi_{H}$ where
\begin{equation}
	\phi_H(t) = \left( \frac{-cH^2}{\lambda} \right)^{1/(n-2)}.
\end{equation}
This is the same order of $\varphi_{\rm min}$ during inflation, and hence $\varphi$ begins oscillation around $\varphi_{\rm min}$
with an oscillation time scale of order one in terms of $s=\ln(t/t_i)$.
Note that the temporal minimum of the potential $\phi_H$ exhibits the same scaling as the scaling solution $\phi \propto t^{-2/(n-2)}$.
Thus $\phi(t)$ follows the temporal minimum $\phi_H(t)$ with oscillation around it.
In this case, this oscillation time scale is close to the Hubble time scale. 
This is qualitatively different from the case of $|c|\ll 1$ in the previous subsection.

Let us see the dynamics more closely.
In order to evaluate the initial amplitude $\Delta\phi$ around the temporal minimum $\phi_H$,
we must specify the origin of the Hubble mass term of $\phi$.
It arises from the Planck-suppressed interaction as
\begin{align}
	-\mathcal L = \frac{\phi^2}{6M_P^2}\left(\frac{c_K}{2}\dot I^2 + c_V V_{\rm inf}(I) \right),
\end{align}
where $c_K$ and $c_V$ are constants of order unity,
$I$ denotes the inflaton filed and $V_{\rm inf}$ is the inflaton potential.
During a slow-roll inflation, the potential energy of the inflaton dominates, hence we have $c=c_V$.
After inflation, the inflaton begins a coherent oscillation and the kinetic energy and potential energy are equally 
distributed, roughly speaking.
In the case of harmonic oscillation of the inflaton, we have $c=(c_K+c_V)/2$.
Therefore, unless $c_K=c_V$, the temporal minimum $\phi_H$ just after inflation is different from that during inflation.
This shift of the potential minimum occurs in the time scale of inflaton mass, rather than the Hubble scale,
so $\phi$ cannot track the temporal minimum~\cite{Nakayama:2011wqa}.
Thus the $\phi$ oscillation around the temporal minimum $\phi_H$ is induced just after inflation,
and the amplitude is of the order of $\Delta\phi \sim \phi_H$.\footnote{
	Even if $c_K=c_V$, $\phi$ cannot track the temporal minimum~\cite{Nakayama:2011wqa}.
}

Another term is coupling to the Ricci scalar $R$:
\begin{equation}
	-\mathcal L = \frac{c_R}{24}\phi^2 R  = \frac{c_R}{2}\phi^2\left(H^2+\frac{\dot H}{2} \right)
	= \frac{c_R \phi^2}{6M_P^2}\left(-\frac{1}{4}\dot I^2 + V_{\rm inf}(I) \right).
\end{equation}
If this is a dominant origin of the Hubble mass term, 
$c=c_R$ during inflation and $c=c_R/4$ after inflation for the harmonic inflaton oscillation.\footnote{
	During the inflaton oscillation, $R$ oscillates between positive and negative values.
}

Anyway, the oscillation amplitude around the temporal minimum $\Delta\phi$ decreases as the universe expands according to
\begin{equation}
	H (\Delta\phi)^2 \propto a^{-3}~~~\leftrightarrow~~~\Delta\phi \propto t^{(1-3p)/2},
\end{equation}
because of the number density conservation.
By using this, we obtain the following scaling:
\begin{align}
	\frac{\Delta\phi}{\phi_H} \sim t^{\frac{2(n_c-n)}{(n-2)(n_c-2)}}.
\end{align}
The oscillation amplitude relatively decreases compared with the position of the temporal minimum for $n>n_c$,
while the ratio is constant for $n=n_c$.
Thus $\phi$ safely relaxes to its temporal minimum for $n>n_c$.
For $n=n_c$, $\phi$ oscillates around $\phi_H$ and never cross $\phi=0$ thereafter if $\Delta\phi \lesssim \phi_H$ initially.
This fact is important to study the formation of topological defects, such as domain walls in the axion model\cite{Harigaya:2015hha}.

%%%%%%%%%%%%%%%%%%%%%%%%%%%%%%%%%%%%%%%%%%%%%%%%%%
\section{Cosmological implications}  \label{sec:cos}
\setcounter{equation}{0}
%%%%%%%%%%%%%%%%%%%%%%%%%%%%%%%%%%%%%%%%%%%%%%%%%%
So far we have considered the dynamics of scalar field with $n=n_c$ having only the
non-renormalizable potential $\phi^n$ and the Hubble mass term.
More realistic models may have other terms in the potential.
In this section we consider cosmological implications of the scalar dynamics with $n=n_c$
based on the potential
\begin{align}
	V = \pm \frac{1}{2}m^2\phi^2 + \frac{\lambda}{n}\phi^n.  \label{pot2}
\end{align}
The presence of the Hubble mass term like $c H^2\phi^2$ does not change the result much as long as $|c| \ll 1$.
In this type of the potential, the dynamics of $\phi$ is eventually dominated by the mass term at $H \lesssim m$,
even if $\phi$ obeys a pseudo scaling law at first.
This is easily seen by rewriting the equation of motion as
\begin{align}
	\varphi'' + \left(3p-\frac{n+2}{n-2}\right)\varphi' +\left(\mu^2\pm\frac{p^2m^2}{H^2}\right)\varphi + \frac{\lambda p^2}{H_i^2} \varphi^{n-1} = 0,  \label{eqofm3}
\end{align}
since $\mu \sim \mathcal O(1)$, the effect of the mass term becomes important at $H \lesssim m$.
Note that even if the potential is dominated by the mass term, the scalar field does not always feel the mass term 
if $\phi$ obeys a pseudo scaling law as we explained at the end of Sec.~\ref{sec:PS}.\footnote{
In fact, $\phi(H=m) \simeq \phi_m (\phi_m/\phi_i) \ll \phi_m$, where $\phi_m$ is defined below Eq.~\eqref{rhom}.
}

\begin{figure}
\begin{center}
\includegraphics[scale = 1.2]{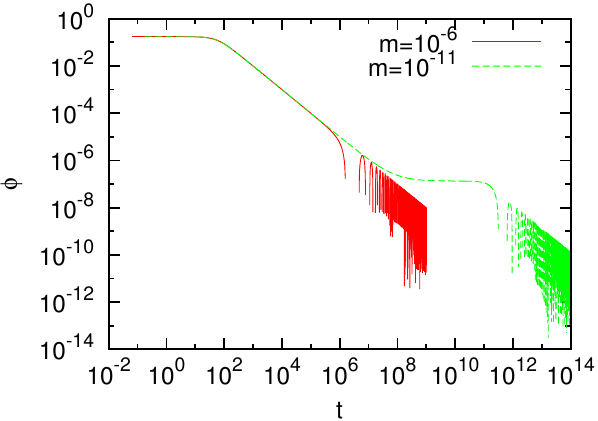}
\includegraphics[scale = 1.2]{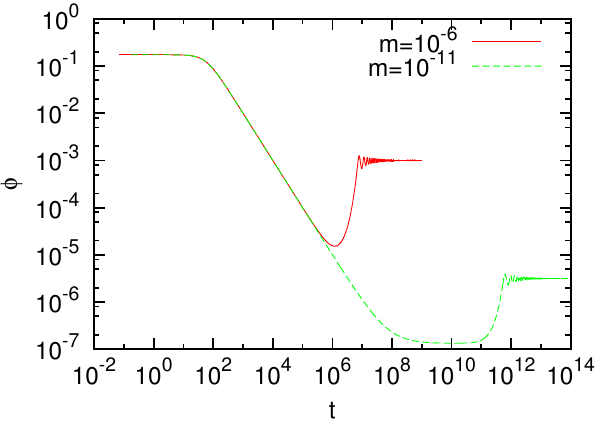}
\end{center}
\caption {
	Time evolution of $\phi(t)$ for $n=6$ and $p=2/3$. 
	In the left (right) panel we have assumed positive (negative) mass squared.
	The red solid (green dashed) line corresponds to $m=10^{-6}$ $(m=10^{-11})$.
}
\label{fig:mass}
\end{figure}

Fig.~\ref{fig:mass} shows numerical results for the time evolution of $\phi(t)$ for $n=6$ for $p=2/3$
with a mass term.
In the left (right) panel we have assumed positive (negative) mass squared.
The red solid (green dashed) line corresponds to $m=10^{-6}$ $(m=10^{-11})$.
Other parameters and initial conditions are the same as those in Fig.~\ref{fig:PS}.
These results clearly show that the pseudo scaling law is maintained even after the mass term dominates the potential
and the oscillation by the mass term starts at at $H \simeq  m$.

%%%%%%%%%%%%%%%%
%%%%%%%%%%%%%%%%

%%%%%%%%%%%%%%%%%%%%%%%%%%%%%%%%%%%%%%%%%%%%%%%%%%
\subsection{Abundance and curvature perturbation}  \label{sec:abu}
%%%%%%%%%%%%%%%%%%%%%%%%%%%%%%%%%%%%%%%%%%%%%%%%%%

First consider the abundance and the curvature perturbation of $\phi$.
Since $\phi$ is light during inflation, it obtains a quantum fluctuation of $\delta\phi_i \simeq H_{\rm inf}/(2\pi)$ 
and $\phi$ may act as a curvaton-like field.
However, it is stated in Ref.~\cite{Dimopoulos:2003ss} that the curvature perturbation vanishes
for both $n > n_c$ and $n=n_c$ because of the scaling behavior of the curvaton field.
If $\phi$ obeys a scaling solution (\ref{scaling}), the field value of $\phi(t)$ at some uniform density slice does not depend on 
the initial position $\phi_i$: this is easily checked as
\begin{equation}
	\phi(t) \simeq \phi_i \left( \frac{t_1}{t} \right)^{\frac{2}{n-2}} \sim \left( \frac{H^2}{\lambda} \right)^{1/(n-2)},
\end{equation}
hence the $\phi_i$ dependence vanishes.
Here $t_1 \sim (\lambda \phi_i^{n-2})^{-1/2}$ is the time at which the effective mass becomes equal to the Hubble scale (see Eq.~(\ref{t1})).
Thus curvature perturbation is not generated by $\phi$ with a scaling law.

This is true for $n>n_c$ in which the scaling solution is an attractor, but for $n=n_c$ we have seen that 
the actual dynamics is far from the scaling law.
For $t_1<t<t_2$, $\phi$ scales as
\begin{equation}
	\phi(t) \simeq \phi_i \left( \frac{t_1}{t} \right)^{4/(n-2)} \sim \frac{1}{\phi_i}\left( \frac{H^2}{\lambda} \right)^{2/(n-2)}.
\end{equation}
The initial $\phi_i$ dependence remains and hence the perturbation $\delta\phi_i$ is not always erased.

For concreteness, let us suppose that $\phi$ decays into radiation at $H = \Gamma_\phi$
with a constant decay rate $\Gamma_\phi$ after the dynamics is dominated by the quadratic mass term for $t_1<t<t_2$ and 
oscillates around the origin.
We also assume that the inflaton energy density scales as $\rho_I \propto a^{-2/p}$ until the curvaton enters in the quadratic oscillation regime,
and decays into radiation before the curvaton decays, i.e., we assume $\Gamma_\phi< \Gamma_I < m$ with $\Gamma_I$
being the inflaton decay rate. 
In this subsection we mainly consider the case of positive mass squared $+m^2$ and give only
results for the case of negative mass squared $-m^{2}$ at the end.
For details on the case of negative mass squared, we refer readers to App.~\ref{sec:hill}.

The energy density of the curvaton at the onset of its oscillation, $H=m$, is evaluated as
\begin{align}
	\rho_{\phi}(H=m) \simeq
	\frac{1}{2}m^2 \phi(H=m)^2 \sim \frac{1}{2}m^2 \phi_m^2\left( \frac{\phi_m}{\phi_i} \right)^2,  \label{rhom}
\end{align}
where we have defined $\phi_m \equiv (m^2/\lambda)^{1/(n-2)}$.
It represents the boundary at which the mass term begins to dominate the potential.
Therefore, the fraction of the $\phi$ energy density at $H=\Gamma_\phi$ is given by
\begin{align}
	\frac{\rho_{\phi}(H=\Gamma_\phi)}{\rho_r(H=\Gamma_\phi)} \simeq
	\frac{1}{6}
	\left( \frac{\phi_m}{M_P} \right)^2
	\left( \frac{\phi_m}{\phi_i} \right)^2
	\left( \frac{\Gamma_I}{\Gamma_\phi} \right)^{1/2}
	\left( \frac{\Gamma_I}{m} \right)^{3p-2}.
\end{align}
This expression assumes that $\phi$ is subdominant at the decay.
Otherwise, it is equal to $1$.

Now, let us evaluate the curvature perturbation generated in this scenario.\footnote{
Here we concentrate on the large scale curvature perturbation.
} From the expression (\ref{rhom}), we can see that the initial $\phi_i$ dependence explicitly remains.
Thus the curvature perturbation of the order of $\zeta \sim \delta\phi_i/\phi_i$ is generated if $\phi$ eventually dominates the universe.

According to the $\delta N$ formalism~\cite{Starobinsky:1986fxa}, the large scale curvature perturbation $\zeta$ is given by
the fluctuation of the e-folding number $N(\vec x)$ between the initial spatially flat surface where the observable scale exit the horizon,
and the final uniform density surface which we can take $H=\Gamma_\phi$.
The total e-folding number can be conveniently divided into $N(\vec x) = N^{(1)}(\vec x)+N^{(2)}(\vec x)+N^{(3)}(\vec x)$ with
\begin{equation}
	N^{(1)} \equiv \ln\left[ \frac{a(H=H_{\rm end})}{a(H=H_{\rm inf})} \right],~~~
	N^{(2)} \equiv \ln\left[ \frac{a(H=\Gamma_I)}{a(H=H_{\rm end})} \right],~~~
	N^{(3)} \equiv \ln\left[ \frac{a(H=\Gamma_\phi)}{a(H=\Gamma_I)} \right],
\end{equation}
where $H_{\rm inf}$ and $H_{\rm end}$ are the Hubble parameters at the horizon crossing and the end of the inflation, respectively.
The curvature perturbation is expanded in terms of the inflaton fluctuation $\delta I_i$ and $\delta\phi_i$ as 
\begin{equation}
	\zeta(\vec x)=\delta N(\vec x) \simeq N_I \delta I_i(\vec x) + N_\phi \delta\phi_i(\vec x) + \frac{1}{2}N_{\phi\phi} (\delta\phi_i (\vec x))^2,
	\label{zeta}
\end{equation}
Note that the inflaton mostly contributes to $N_1$ and $\phi$ can have its largest contribution to $N_3$,
hence we can safely approximate as $ N_I \simeq N^{(1)}_I $, $N_\phi \simeq N^{(3)}_\phi$ and $N_{\phi\phi} \simeq N^{(3)}_{\phi\phi}$.
Thus we immediately have
\begin{equation}
	N^{(1)}_I= \frac{V_{\rm inf}}{M_P^2( \partial_{I_i}V_{\rm inf})}.
\end{equation}
To derive $N^{(3)}_\phi$ and $N^{(3)}_{\phi\phi}$, we use the relation
at the curvaton decay surface $H(\vec x)=\Gamma_\phi$~\cite{Sasaki:2006kq}, 
\begin{equation}
	\rho_r(H=\Gamma_\phi, \vec x) + \rho_\phi(H=\Gamma_\phi,\vec x)= 3\Gamma_\phi^2M_P^2,
\end{equation}
where $\rho_r$ denotes the radiation energy density that is created by the inflaton decay.
They are expressed by using the e-foldings of the scale factor
\begin{align}
	\rho_r(H=\Gamma_\phi, \vec x) &= \rho_I(H=\Gamma_I)e^{-4N^{(3)}(\vec x)},\\
	\rho_\phi(H=\Gamma_\phi, \vec x) &=\rho_\phi(H=\Gamma_I,\phi_i(\vec x))e^{-3N^{(3)}(\vec x)}.
\end{align}
By differentiating this expression with respect to $\phi_i$, we obtain
\begin{align}
	N^{(3)}_\phi = r\frac{\partial_{\phi_i}\rho_\phi}{3\rho_\phi},~~~
	N^{(3)}_{\phi\phi}=r\left[ \frac{\partial^2_{\phi_i}\rho_\phi}{3\rho_\phi} -(r^2+2r)\left(\frac{\partial_{\phi_i}\rho_\phi}{3\rho_\phi}\right)^2  \right].
\end{align}
where
\begin{align}
	r &\equiv \left[\frac{3 \bar\rho_\phi }{4\bar\rho_r + 3\bar\rho_\phi}\right]_{H=\Gamma_\phi},
\end{align}
with the over-lines indicating the averaged quantities.

In the case of positive mass squared $+m^{2}$, to the second order in $\delta\phi$, we obtain the curvature perturbation as
\begin{align}
	\zeta(\vec x)=-\frac{2r}{3}\frac{\delta\phi_i(\vec x)}{\phi_i}
	+ r\left[ 1- \frac{2(2r+r^2)}{9} \right]\left(\frac{\delta\phi_i(\vec x)}{\phi_i} \right)^2,
\end{align}
where we have neglected the inflaton fluctuation.
The non-linearity parameter of the local type is given by,
\begin{equation}
	\frac{6}{5} f_{\rm NL} = \frac{9}{2r}\left[ 1- \frac{2(2r+r^2)}{9} \right] \frac{N_\phi^4}{(N_I^2+N_\phi^2)^2}.
\end{equation}

Here we also show the results in the case of negative mass squared $-m^{2}$.\footnote{
The small scale perturbations do not grow even in the presence of the tachyonic mass
if the momentum is larger than the tachyonic mass.
} If the mass term dominates 
the dynamics during the regime $t_{2} < t < t_{3}$, the curvature perturbation 
is given by
\begin{align}
	\zeta = N_I \delta I + (n+1)pr \beta\left[ -\frac{\delta\phi_i}{\phi_i} + \frac{1}{2}\left(\frac{\delta\phi_i}{\phi_i} \right)^2\right],
\end{align}
where $\beta \sim \mathcal{O}(10^{-1})$ is a numerical factor. 
The non-linearity parameter is given by
\begin{align}
	\frac{6}{5}f_{\rm NL} =  \frac{1}{(n+1)pr\beta} \frac{N_\phi^4}{(N_I^2+N_\phi^2)^2}.
\end{align}
For more details, see App.~\ref{sec:hill}.

This is one example of the property of the curvature perturbation from the pseudo scaling curvaton.
Although these expressions depend on how the background evolves and when the curvaton decays,
it generally holds that the curvature perturbation is generated for $n=n_c$ in which $\phi$ obeys a pseudo scaling law.

%%%%%%%%%%%%%%%%%%%%%%%%%%%%%%%%%%%%%%%%%%%%%%%%%%
\subsection{Domain wall problem}  \label{sec:wall}
%%%%%%%%%%%%%%%%%%%%%%%%%%%%%%%%%%%%%%%%%%%%%%%%%%

Next, we consider implications on the domain wall formation.
Consider the case of negative mass squared: $-m^2$. The potential 
has a $Z_2$ symmetry $\phi \rightarrow -\phi$ if $n$ is even.
If $\phi$ obtains a large initial value during inflation and $n < n_{c}$,
it oscillates around the symmetric point $\phi = 0$ 
after inflation and large fluctuations of $\phi$ particles (and other particles 
which interact with $\phi$) are induced due to the efficient particle production.
Such large fluctuations may cause the non-thermal restoration of the symmetry 
and $\phi$ may be trapped at the origin ($\phi = 0$)~\cite{Kofman:1995fi,Tkachev:1995md}. 
The trapping effect becomes diluted as the universe expands, and $\phi$ finally rolls down to the minimum
of the potential. Eventually, the domain wall is formed and the universe is
expected to be overclosed. This is the rough picture of the domain wall problem we consider here.
A physically well-motivated example is the Peccei-Quinn (PQ) field, which often leads to the axionic domain wall problem~\cite{Kawasaki:2013ae}
due to the nonthermal restoration of the PQ symmetry~\cite{Kasuya:1996ns, Moroi:2013tea}.

On the other hand, for $n = n_{c}$, the scalar field $\phi$ obeys the pseudo scaling law.
As we saw earlier, the energy density of $\varphi$ is negative for a reasonable initial condition
(see Eq.~\eqref{eq:energy}). Hence $\phi$ never crosses the origin during the pseudo scaling
and no particle production occurs.\footnote{
Even if $\rho_{\varphi} > 0$, no efficient particle production is expected to occur since the 
period of the oscillation is much longer than the cosmic time in the case of the pseudo scaling solution.} 
Moreover, even after the dynamics is dominated by the mass term, $\phi$ never crosses
the origin because the mass term also prevents $\varphi$ from crossing the origin (see Eq.~\eqref{eqofm3}).
The field $\phi$ just oscillates around $\phi_{m}$ and the domain wall problem does not occur. Therefore, the 
pseudo scaling solution provides a natural solution to the domain wall problem.

Here we emphasize that the pseudo scaling solution for $n = n_{c}$ is more powerful 
in solving the domain wall problem than the scaling solution for $n > n_{c}$.
To see this point, let us consider the following potential:
\begin{align}
V = -\frac{m^{2}}{2}\phi^{2} + \frac{\lambda_l}{l}\phi^l + \frac{\lambda}{n}\phi^{n},
\end{align}
where we assume $n, n_{c} > l$ and $l > 2$. First consider the scaling solution $n > n_{c}$. In this case,
if the initial value of $\phi$ is large enough, the field $\phi$ obeys the scaling solution
after inflation. If $\phi_{l} \equiv (\lambda_l/\lambda)^{1/(n-l)} \gg \phi_{m}$, the dynamics
is dominated by the $\phi^{l}$ term and $\phi$ starts to oscillate around the origin
when $\phi \simeq \phi_{l}$. Thus, for sufficiently large $\phi_l$, the domain wall problem 
occurs even if $\phi$ obeys the scaling solution at first. 

On the other hand, in the 
case of the pseudo scaling solution $n = n_{c}$, the dynamics becomes dominated by
the $\phi^{l}$ term when $H \simeq H_{l}$ where $H_{l}$ is defined as\footnote{
Note that $H_{l}/m = (\phi_{l}/\phi_{m})^{(n-2)/2}$, and hence the dynamics is dominated by the $\phi^{l}$
term before the mass term becomes effective for $\phi_{l} \gg \phi_{m}$.
}
\begin{align}
\frac{H_{l}}{\sqrt{\lambda}} \equiv \left(\frac{\lambda_{l}}{\lambda}\right)^{\frac{n-2}{2(n-l)}} = \phi_{l}^{\frac{n-2}{2}}.
\label{eq:h4}
\end{align}
We can deduce this as follows. In terms of $\varphi$, the effective potential is given by
\begin{align}
V_{\rm eff}(\varphi) = \frac{1}{2}\left(\mu^{2} - \frac{p^{2}m^{2}}{H^{2}}\right)\varphi^{2}
+ \frac{\lambda_{l}p^{2}}{lH_{i}^{2}}\left(\frac{H_{i}}{H}\right)^{\frac{2(n-l)}{n-2}}\varphi^{l} + \frac{\lambda p^{2}}{nH_{i}^{2}}\varphi^{n}.
\end{align}
The amplitude of $\varphi$ is of the order of $\varphi_{\rm min}$, and hence the $\varphi^{l}$ term becomes effective
once the condition
\begin{align}
\frac{\lambda_{l}}{H_{i}^{2}}\left(\frac{H_{i}}{H}\right)^{\frac{2(n-l)}{n-2}}\varphi_{\rm min}^{2} \sim \mathcal{O}(1),
\end{align}
is satisfied. This condition is equivalent to Eq.~\eqref{eq:h4}.
Therefore, for example, if the $\phi^{l}$ term dominates the dynamics for the regime $t_{1} < t < t_{2}$, 
the field value of $\phi$ at $H=H_{l}$ is given by
\begin{align}
\phi(H=H_{l}) \simeq \left(\frac{\phi_{l}}{\phi_{i}}\right)\phi_{l}.
\end{align}
If the initial amplitude satisfies $\phi_{i} \gg \phi_{l}$, $\phi$ starts to oscillate
with the amplitude much smaller than $\phi_{l}$. Thus, the domain wall problem is less likely to occur even in 
this case compared to the scaling solution.

So far we have focused only on the domain wall problem, but similar implications can be obtained for
other types of topological defects.

%%%%%%%%%%%%%%%%%%%%%%%%%%%%%%%%%%%%%%%%%%%%%%%%%%
\section{Conclusions and discussion}
\label{sec:conclusion}
\setcounter{equation}{0}
%%%%%%%%%%%%%%%%%%%%%%%%%%%%%%%%%%%%%%%%%%%%%%%%%%

We have shown that a scalar field with potential $\phi^n$ with $n=n_c$ exhibits a peculiar behavior
which we call a pseudo scaling.
It resembles neither oscillating nor scaling solution.
We have also shown that the primordial curvature perturbation is generated by this pseudo-scaling curvaton.
The critical exponent $n_c$ depends on the background expansion law:
$n_c=6$ $(n_c=10)$ in the MD (RD) universe.
Even a renormalizable term can lead to a pseudo-scaling dynamics, i.e., $n\leq 4$ is obtained for $p\geq 1$,
where the Hubble parameter is given by $H = p/t$.

This observation has implications on some realistic models and cosmological scenarios.
In supersymmetric (SUSY) theories, for example, it is often the case that a scalar field has a non-renormalizable potential.
An example is the Affleck-Dine (AD) field~\cite{Affleck:1984fy} which consists of the flat direction in the MSSM and 
have the superpotential of the form~\cite{Dine:1995kz,Gherghetta:1995dv}
\begin{equation}
	W = \frac{\gamma \phi^\ell}{\ell},
\end{equation}
where $\phi$ denotes the AD field and the scalar potential is then given by
\begin{align}
	V = m^2|\phi|^2 + A_m\left( \frac{\gamma\phi^\ell}{\ell}+{\rm h.c.} \right)+ \gamma^2|\phi|^{2(\ell-1)},
\end{align}
where $m$ and $A_m$ represent the effect of SUSY breaking.
If there is no Hubble mass term or its coefficient is small, and the last term dominates the potential,
the AD field may enter the pseudo-scaling regime.
The $LH_u$ flat direction, which is lifted up by the $\ell=4$ superpotential, matches the condition for the pseudo scaling
if the background is matter-dominated as in the inflaton oscillation dominated era.
In such a case, the lepton or baryon asymmetry generated by the AD mechanism will be significantly suppressed 
compared with the naive expectation, e.g., the case of negative Hubble mass term.

Another example is the SUSY axion model, in which the superpotential is given by~\cite{Murayama:1992dj}
\begin{align}
	W = \frac{\gamma \phi^\ell \bar\phi}{\ell},
\end{align}
where $\phi$ and $\bar\phi$ are the PQ field having PQ charge $+1$ and $-\ell$ respectively. The scalar potential is
\begin{align}
	V = -m^2|\phi|^2 + \gamma^2|\phi|^{2\ell},
\end{align}
where we have taken $\bar\phi=0$ for simplicity.
If $m^2>0$, $\phi$ can have a large VEV which may be suitable as a solution to the strong CP problem~\cite{Peccei:1977hh,Kim:1986ax}.

In the case where the PQ symmetry is broken before inflation, the axion field usually obtains a large fluctuation and may 
cause the isocurvature perturbation problem. One way to avoid the isocurvature problem is to consider the situation where
the PQ field takes a very large value during inflation~\cite{Linde:1990yj}. In such a case, the isocurvature perturbation gets suppressed due to the 
large expectation value of the PQ field. However, since the initial field value is large, particles are expected to
be effectively produced due to the oscillation of the PQ field after inflation. As a result, the PQ symmetry may be
non-thermally restored and hence the domain wall problem may occur after the QCD phase 
transition~\cite{Kasuya:1996ns,Moroi:2013tea}. 
In this paper, we have seen that if the PQ scalar obeys the pseudo scaling law after inflation, no such a disastrous situation is realized.
Thus, for example, if $l = 3$ and the PQ scalar starts to move during the inflaton oscillation dominated era (with the mass term),
the domain wall problem as well as the isocurvature perturbation problem can be naturally solved.

%%%%%%%%%%%%%%%%%%%%%%%%%%%%%%%%%%%%%%%%%%%%
\section*{Acknowledgments}
%%%%%%%%%%%%%%%%%%%%%%%%%%%%%%%%%%%%%%%%%%%%

This work was supported by the Grant-in-Aid for Scientific Research on Scientific Research A (No.26247042 [KN]),
Young Scientists B (No.26800121 [KN]) and Innovative Areas (No.26104009 [KN], No.15H05888 [KN]).
This work was supported by World Premier International Research Center Initiative (WPI Initiative), MEXT, Japan. 
The work of Y.E. and M.T. was supported in part by JSPS Research Fellowships for Young Scientists 
and the Program for Leading Graduate Schools, MEXT, Japan.

\appendix

%%%%%%%%%%%%%%%%%%%%%%%%%%%%%%%%%%%%%%%%%%%%
\section{Expansion law in the inflaton oscillation dominated era}  \label{sec:app}
%%%%%%%%%%%%%%%%%%%%%%%%%%%%%%%%%%%%%%%%%%%%

In this appendix we derive the expansion law $a\propto t^p$ or $H=p/t$ in the inflaton oscillation dominated universe.
We assume that the inflaton potential around the potential minimum is approximated as 
\begin{equation}
	V_{\rm inf}(I) \propto I^q,
\end{equation}
where $I$ denotes the inflaton field measured from its potential minimum.
The equation of motion of the inflaton field is given by
\begin{equation}
	\ddot I + 3H \dot I + \frac{\partial V_{\rm inf}}{\partial I} = 0.
\end{equation}
From this we obtain
\begin{equation}
	\dot \rho_I + 3H \dot I^2 = 0,
\end{equation}
where $\rho_I = \dot I^2/2 + V_{\rm inf}(I)$ is the inflaton energy density.
On the other hand, we have
\begin{equation}
	\frac{d}{dt}(I\dot I) = \dot I^2 - qV_{\rm inf} - 3HI\dot I.
\end{equation}
Taking time average over the inflaton oscillation time period and assuming that the Hubble parameter is much smaller than the
inflaton mass scale, we obtain the Virial theorem 
\begin{equation}
	\langle \dot I^2 \rangle = q\langle V_{\rm inf} \rangle,
\end{equation}
where $\langle\cdots\rangle$ indicates time average over the inflaton oscillation period.
By using this relation, we obtain
\begin{align}
	\dot\rho_I + \frac{6q}{q+2}H \rho_I = 0.
\end{align}
Thus the energy density scales as
\begin{equation}
	\rho_I \propto a^{-\frac{6q}{q+2}}.
\end{equation}
On the  other hand, the Freedman equation $3H^2M_P^2=\rho_I$ implies $\rho_I \propto t^{-2}$.
Therefore, we obtain the expansion law
\begin{equation}
	a \propto t^p,~~~p=\frac{q+2}{3q}.  \label{pinf}
\end{equation}
For $q=2$ $(q=4)$, the usual MD (RD)-like universe $p=2/3$ $(p=1/2)$ is realized.
More unconventional expansion laws are also obtained, for example, $p=1$ for $q=1$ and $p=5/3$ for $q=1/2$.
In terms of $q$, the critical exponent $n_{c}$ is given by
\begin{align}
n_{c} = 2(q+1).
\end{align}
%%

%%%%%%%%%%%%%%%%%%%%%%%%%%%%%%%%%%%%%%%%%%%%
\section{Formal solution of a pseudo scaling scalar field}
\label{sec:app2}
\setcounter{equation}{0}
%%%%%%%%%%%%%%%%%%%%%%%%%%%%%%%%%%%%%%%%%%%%
In this appendix, we give a formal solution of the equation of motion~\eqref{eqofm2}.
Note that the energy density of $\varphi$
\begin{align}
\rho_{\varphi} = \frac{1}{2}\varphi^{\prime 2} + V_{\rm eff}(\varphi),
\end{align}
is conserved. Thus, the kinetic term of $\varphi$ is given by
\begin{align}
\varphi^{\prime} = \pm\sqrt{2\left(\rho_{\varphi} - V_{\rm eff}(\varphi)\right)}.
\end{align}
By integrating this, we obtain the formal solution of $\varphi$ as
\begin{align}
s - s_{i} = \frac{1}{\sqrt{2}}\int_{\varphi_{i}}^{\varphi}\frac{d\varphi}{\sqrt{\rho_{\varphi} - V_{\rm eff}}},
\end{align}
for $\varphi^{\prime} > 0$ region and
\begin{align}
s - s_{i} = \frac{1}{\sqrt{2}}\int^{\varphi_{i}}_{\varphi}\frac{d\varphi}{\sqrt{\rho_{\varphi} - V_{\rm eff}}},
\end{align}
for $\varphi^{\prime} < 0$ region. Here $s_{i}$ and $\varphi_{i}$ are some initial values
of $s$ and $\varphi$, respectively. In particular, the period of the oscillation of $\varphi$
is given by
\begin{align}
s_{\rm osc} = \sqrt{2}\int_{\varphi_{-}}^{\varphi_{+}}\frac{d\varphi}{\sqrt{\rho_{\varphi} - V_{\rm{eff}}}},
\end{align}
where $\varphi_{\pm}$ are the solutions of the equation $\rho_{\varphi} = V_{\rm eff}(\varphi_{\pm})$ with $\varphi_{+} > \varphi_{-}$.\footnote{
Here we assume that $\rho_{\varphi} < 0$. For $\rho_{\varphi} > 0$, we must replace $\varphi_{-}$ by zero and 
also double the right hand side of the equation when $n$ is even.
} Note that $\varphi_{\pm}$ as well as $\rho_{\varphi}$ depends on the initial condition.
However, it is difficult to solve this equation analytically. Moreover, 
the approximation given in Sec.~\ref{sec:PS} is sufficient for our purpose in this paper.

%%%%%%%%%%%%%%%%%%%%%%%%%%%%%%%%%%%%%%%%%%%%
\section{Curvaton with hilltop potential}  \label{sec:hill}
\setcounter{equation}{0}
%%%%%%%%%%%%%%%%%%%%%%%%%%%%%%%%%%%%%%%%%%%%

In this section we briefly review the curvature perturbation generated by the curvaton with a 
hilltop potential of the form
\begin{align}
	V(\phi) = -\frac{1}{2}m^2\phi^2 + \frac{\lambda}{n}\phi^n.  \label{tachyon}
\end{align}
As shown in Sec.~\ref{sec:cos}, if the initial position $\phi_i$ is set $\phi_m \ll \phi_i \lesssim \varphi_{\rm min}$,
where $\phi_m\equiv (m^2/\lambda)^{1/(n-2)}$ is the minimum position of the potential and $\varphi_{\rm min} \sim (H_i^2/\lambda)^{1/(n-2)}$,
$\phi$ obeys a pseudo scaling law until $H$ becomes comparable to $m$.
For concreteness, let us suppose that this breaking of pseudo scaling law happens at $t_2 < t < t_3$.
The field value at this stage is given by (\ref{phit2}):
\begin{equation}
	\phi_* = \frac{\phi_i^{n+1}}{\varphi_{\rm min}^n},  \label{phi_ini}
\end{equation}
and this can be much smaller than $\phi_m$.
The scalar dynamics thereafter is the same as that with a scalar potential (\ref{tachyon}) with initial position $\phi_*$ close to the origin.
In other words, the pseudo scaling scalar dynamics naturally provides us with a hilltop initial condition.
In the following, we treat $\phi_* (\ll \phi_m)$ as if it is the ``initial'' condition of the dynamics,
although the true initial position is $\phi_i$.\footnote{
	Here and in what follows we focus on the region $\phi > 0$ without loss of generality.
}

As in Sec.~\ref{sec:abu}, let us assume $\Gamma_\phi < \Gamma_I < m$.
The abundance of the curvaton at $H=\Gamma_\phi$ is evaluated as
\begin{align}
	\frac{\rho_\phi(H=\Gamma_\phi)}{\rho_r(H=\Gamma_\phi)}
	\simeq \frac{n-2}{6}\left( \frac{\phi_m-\phi_{\rm os}}{M_P} \right)^2
	\left( \frac{m}{H_{\rm os}} \right)^2\left( \frac{\Gamma_I}{\Gamma_\phi} \right)^{1/2}
	\left( \frac{\Gamma_I}{H_{\rm os}} \right)^{3p-2}.
\end{align}
This expression assumes that $\phi$ is subdominant at the decay. Otherwise, it is equal to 1.
Here $H_{\rm os}$ and $\phi_{\rm os}$ denote the Hubble parameter and the curvaton field value at the onset of its oscillation, respectively.
What is nontrivial in the hilltop scenario is that $H_{\rm os}$ strongly depends on the initial condition $\phi_*$ in the hilltop limit $\phi_*\ll \phi_m$\cite{Kawasaki:2008mc,Kawasaki:2011pd}.

Let us express $H_{\rm os}$ and $\phi_{\rm os}$ in terms of $\phi_*$.
Before the onset of oscillation, we can neglect the second term of (\ref{tachyon}).
The formal solution of the equation of motion with a tachyonic mass term $-m^2\phi^2/2$ and $H=p/t$ is written as
\begin{equation}
	\phi(t) =  \phi_* 2^\nu\Gamma(\nu+1) (mt)^{-\nu} I_\nu(mt),~~~~~\nu\equiv\frac{3p-1}{2},
\end{equation}
where $I_\nu(x)$ is the modified Bessel function of the first kind.
In general, the solution is a superposition of the modified Bessel function of the first and second kind.
However, without loss of generality we can consider the region $\phi>0$ and $\dot\phi>0$, 
and hence the Bessel function of the first kind is sufficient to describe the dynamics.
The solution is conveniently approximated as
\begin{equation}
	\phi(t) \simeq \begin{cases}
	 \displaystyle \phi_* \left[1+ \frac{(mt)^2}{2(3p+1)} \right]  &~~~{\rm for}~~~mt \ll 1,\\
	 \displaystyle \phi_* \frac{2^\nu\Gamma(\nu+1)}{\sqrt{2\pi}} \frac{e^{mt}}{(mt)^{3p/2}}&~~~{\rm for}~~~mt \gg 1.
	\end{cases}
	\label{phi_sol}
\end{equation}
First note that the velocity of the scalar field is given by\footnote{
	For $|V''/H^2|\ll 1$, we have a solution for a general potential: $H\dot\phi= -pV'/(3p+1)$.
}
\begin{align}
	\dot\phi(t) \sim \begin{cases}
	\displaystyle\frac{p}{3p+1}\frac{m^2}{H}\phi&~~~{\rm for}~~~mt \ll 1,  \\
	\displaystyle m \phi  &~~~{\rm for}~~~mt \gg 1.
	\end{cases}
	 \label{dotphi}
\end{align}
Since the oscillation epoch is significantly delayed in the hilltop limit $\phi_*\ll \phi_m$, i.e., $m/H_{\rm os} \gg 1$,
the condition of the start of the oscillation is given by
\begin{equation}
	\left| \frac{\dot\phi}{H\phi_m} \right|  \simeq 1
	~~~\leftrightarrow~~~ H_{\rm os} \simeq \frac{m \phi_{\rm os}}{\phi_m}.
\end{equation}
On the other hand, from Eq.~(\ref{phi_sol}), we obtain
\begin{equation}
	\phi_{\rm os} \simeq \phi_m\frac{1}{\ln(\phi_{\rm os}/\phi_*)}  \equiv \beta \phi_m.
\end{equation}
For reasonable choice of $\phi_* \ll \phi_m$, we have $\beta \sim \mathcal O(10^{-1})$.
This justifies our assumption $H_{\rm os} \ll m$.
Then following relations are derived,
\begin{align}
	\frac{\partial \phi_{\rm os}}{\partial \phi_*} \simeq \frac{\beta}{1+\beta} \frac{\phi_{\rm os}}{\phi_*},~~~~~
	\frac{\partial^2 \phi_{\rm os}}{\partial \phi_*^2} \simeq -\frac{\beta}{(1+\beta)^3} \frac{\phi_{\rm os}}{\phi_*^2}.
\end{align}

Using these equations and noting that $\rho_\phi(H=\Gamma_\phi) \propto H_{\rm os}^{-3p}$, we obtain 
\begin{align}
	\frac{\partial_{\phi_*}\rho_\phi}{\rho_\phi}\simeq -\frac{3p\beta}{\phi_*},~~~~~~
	\frac{\partial^2_{\phi_*}\rho_\phi}{\rho_\phi}\simeq \frac{3p\beta}{\phi_*^2},
\end{align}
where we have taken only leading terms in $\beta$.
Hence the curvature perturbation (\ref{zeta}) is given by
\begin{align}
	\zeta = N_I \delta I + pr \beta\left[ -\frac{\delta\phi_*}{\phi_*} + \frac{1}{2}\left(\frac{\delta\phi_*}{\phi_*} \right)^2\right],
\end{align}
to the leading order in $\beta$.
If $\phi_*$ would be a true initial condition set during inflation, then $\delta \phi_* \sim H_{\rm inf}/(2\pi)$
and this expression gives a correct estimate for the curvature perturbation from the hilltop curvaton.
In this case, the non-linearity parameter of $f_{\rm NL} = 5/(6pr\beta) \sim \mathcal O(10)$ for $r\simeq 1$
is a generic prediction\cite{Kawasaki:2011pd}.

In our pseudo-scaling curvaton case, the true initial position is $\phi_i$
and it is $\delta\phi_i$ that obtains a quantum fluctuation of $\sim H_{\rm inf}/(2\pi)$.
From the relation (\ref{phi_ini}), we obtain 
\begin{align}
	\zeta = N_I \delta I + (n+1)pr \beta\left[ -\frac{\delta\phi_i}{\phi_i} + \frac{1}{2}\left(\frac{\delta\phi_i}{\phi_i} \right)^2\right]. 
\end{align}
The non-linearity parameter is given by
\begin{align}
	\frac{6}{5}f_{\rm NL} =  \frac{1}{(n+1)pr\beta} \frac{N_\phi^4}{(N_I^2+N_\phi^2)^2}.
\end{align}
Therefore, if the inflaton contribution to the curvature perturbation is negligible, 
we have $f_{\rm NL} \sim \beta^{-1}/n \sim \mathcal{O}(1\mathchar`-10)$ for $r\simeq 1$.
Thus in this case the non-linearity parameter may become sizable.

%%%%%%%%%%%%%%%%%%%%%%%%%%%%%%%%%%%%%%%%%%%%%%%%%%

%%%%%%%%%%%%%%%%%%%%%%%%%%%%%%%%%%%%%%%%%%%%%%%%%%


\begin{thebibliography}{99}
%%%%%%%%%%%%%%%%%%%%%%%%%%%%%%%%%%%%%%%%%%%%%%%%%%

   %\cite{Guth:1980zm}
\bibitem{Guth:1980zm}
  A.~H.~Guth,
  %``The Inflationary Universe: A Possible Solution to the Horizon and Flatness Problems,''
  Phys.\ Rev.\  D {\bf 23}, 347-356 (1981);
  %\cite{Starobinsky:1980te}
%\bibitem{Starobinsky:1980te}
A.~A.~Starobinsky,
%``A New Type of Isotropic Cosmological Models without Singularity,''
Phys.\ Lett.\ B {\bf 91} (1980) 99;
%%CITATION = PHLTA,B91,99;%%
%\cite{Sato:1980yn}
%\bibitem{Sato:1980yn}
  K.~Sato,
  %``First Order Phase Transition of a Vacuum and Expansion of the Universe,''
  Mon.\ Not.\ Roy.\ Astron.\ Soc.\  {\bf 195}, 467-479 (1981);  
  %\cite{Linde:1981mu}
%\bibitem{Linde:1981mu}
A.~D.~Linde,
%``A New Inflationary Universe Scenario: a Possible Solution of the Horizon, Flatness, Homogeneity, Isotropy and Primordial Monopole Problems,''
Phys.\ Lett.\ B {\bf 108} (1982) 389;
%%CITATION = PHLTA,B108,389;%%
%\cite{Albrecht:1982wi}
%\bibitem{Albrecht:1982wi} 
  A.~Albrecht and P.~J.~Steinhardt,
  %``Cosmology for Grand Unified Theories with Radiatively Induced Symmetry Breaking,''
  Phys.\ Rev.\ Lett.\  {\bf 48}, 1220 (1982).
  %%CITATION = PRLTA,48,1220;%%

%\cite{Turner:1983he}
\bibitem{Turner:1983he} 
  M.~S.~Turner,
  %``Coherent Scalar Field Oscillations in an Expanding Universe,''
  Phys.\ Rev.\ D {\bf 28}, 1243 (1983).
  %%CITATION = PHRVA,D28,1243;%%
  
  %\cite{Mollerach:1989hu}
\bibitem{Mollerach:1989hu} 
  S.~Mollerach,
  %``Isocurvature Baryon Perturbations and Inflation,''
  Phys.\ Rev.\ D {\bf 42}, 313 (1990);
  %%CITATION = PHRVA,D42,313;%%
  %\cite{Linde:1996gt}
%\bibitem{Linde:1996gt} 
  A.~D.~Linde and V.~F.~Mukhanov,
  %``Nongaussian isocurvature perturbations from inflation,''
  Phys.\ Rev.\ D {\bf 56}, 535 (1997)
  [astro-ph/9610219];
  %%CITATION = ASTRO-PH/9610219;%%
%\cite{Enqvist:2001zp}
%\bibitem{Enqvist:2001zp} 
  K.~Enqvist and M.~S.~Sloth,
  %``Adiabatic CMB perturbations in pre - big bang string cosmology,''
  Nucl.\ Phys.\ B {\bf 626}, 395 (2002)
  [hep-ph/0109214];
  %%CITATION = HEP-PH/0109214;%%
  %\cite{Lyth:2001nq}
%\bibitem{Lyth:2001nq} 
  D.~H.~Lyth and D.~Wands,
  %``Generating the curvature perturbation without an inflaton,''
  Phys.\ Lett.\ B {\bf 524}, 5 (2002)
  [hep-ph/0110002];
  %%CITATION = HEP-PH/0110002;%%
%\cite{Moroi:2001ct}
%\bibitem{Moroi:2001ct} 
  T.~Moroi and T.~Takahashi,
  %``Effects of cosmological moduli fields on cosmic microwave background,''
  Phys.\ Lett.\ B {\bf 522}, 215 (2001)
  [Erratum-ibid.\ B {\bf 539}, 303 (2002)]
  [hep-ph/0110096].
  %%CITATION = HEP-PH/0110096;%%
  
%\cite{Dimopoulos:2003ss}
\bibitem{Dimopoulos:2003ss} 
  K.~Dimopoulos, G.~Lazarides, D.~Lyth and R.~Ruiz de Austri,
  %``Curvaton dynamics,''
  Phys.\ Rev.\ D {\bf 68}, 123515 (2003)
  [hep-ph/0308015].
  %%CITATION = HEP-PH/0308015;%%
  
  %\cite{Enqvist:2009zf}
\bibitem{Enqvist:2009zf} 
  K.~Enqvist, S.~Nurmi, G.~Rigopoulos, O.~Taanila and T.~Takahashi,
  %``The Subdominant Curvaton,''
  JCAP {\bf 0911}, 003 (2009)
  [arXiv:0906.3126 [astro-ph.CO]];
  %%CITATION = ARXIV:0906.3126;%%
  %\cite{Enqvist:2009ww}
%\bibitem{Enqvist:2009ww} 
  K.~Enqvist, S.~Nurmi, O.~Taanila and T.~Takahashi,
  %``Non-Gaussian Fingerprints of Self-Interacting Curvaton,''
  JCAP {\bf 1004}, 009 (2010)
  [arXiv:0912.4657 [astro-ph.CO]].
  %%CITATION = ARXIV:0912.4657;%%
  
  %\cite{Byrnes:2011gh}
\bibitem{Byrnes:2011gh} 
  C.~T.~Byrnes, K.~Enqvist, S.~Nurmi and T.~Takahashi,
  %``Strongly scale-dependent polyspectra from curvaton self-interactions,''
  JCAP {\bf 1111}, 011 (2011)
  [arXiv:1108.2708 [astro-ph.CO]].
  %%CITATION = ARXIV:1108.2708;%%
  
  %\cite{Mukaida:2014wma}
\bibitem{Mukaida:2014wma} 
  K.~Mukaida, K.~Nakayama and M.~Takimoto,
  %``Suppressed Non-Gaussianity in the Curvaton Model,''
  Phys.\ Rev.\ D {\bf 89}, no. 12, 123515 (2014)
  [arXiv:1402.1856 [astro-ph.CO]].
  %%CITATION = ARXIV:1402.1856;%%
  
  %\cite{Liddle:1998xm}
\bibitem{Liddle:1998xm} 
  A.~R.~Liddle and R.~J.~Scherrer,
  %``A Classification of scalar field potentials with cosmological scaling solutions,''
  Phys.\ Rev.\ D {\bf 59}, 023509 (1999)
  [astro-ph/9809272].
  %%CITATION = ASTRO-PH/9809272;%%
  
     %\cite{Peccei:1977hh}
\bibitem{Peccei:1977hh} 
  R.~D.~Peccei and H.~R.~Quinn,
  %``CP Conservation in the Presence of Instantons,''
  Phys.\ Rev.\ Lett.\  {\bf 38}, 1440 (1977).
  %%CITATION = PRLTA,38,1440;%%
  
   %\cite{Affleck:1984fy}
\bibitem{Affleck:1984fy} 
  I.~Affleck and M.~Dine,
  %``A New Mechanism for Baryogenesis,''
  Nucl.\ Phys.\ B {\bf 249}, 361 (1985).
  %%CITATION = NUPHA,B249,361;%%
  
  %\cite{Dine:1995kz}
\bibitem{Dine:1995kz} 
  M.~Dine, L.~Randall and S.~D.~Thomas,
  %``Baryogenesis from flat directions of the supersymmetric standard model,''
  Nucl.\ Phys.\ B {\bf 458}, 291 (1996)
  [hep-ph/9507453].
  %%CITATION = HEP-PH/9507453;%%
  
    %\cite{Harigaya:2015hha}
\bibitem{Harigaya:2015hha} 
  K.~Harigaya, M.~Ibe, M.~Kawasaki and T.~T.~Yanagida,
  %``Dynamics of Peccei-Quinn Breaking Field after Inflation and Axion Isocurvature Perturbations,''
  arXiv:1507.00119 [hep-ph].
  %%CITATION = ARXIV:1507.00119;%%
  
  %\cite{Nakayama:2011wqa}
\bibitem{Nakayama:2011wqa} 
  K.~Nakayama, F.~Takahashi and T.~T.~Yanagida,
  %``On the Adiabatic Solution to the Polonyi/Moduli Problem,''
  Phys.\ Rev.\ D {\bf 84}, 123523 (2011)
  [arXiv:1109.2073 [hep-ph]].
  %%CITATION = ARXIV:1109.2073;%%
      
  %\cite{Starobinsky:1986fxa}
\bibitem{Starobinsky:1986fxa} 
  A.~A.~Starobinsky,
  %``Multicomponent de Sitter (Inflationary) Stages and the Generation of Perturbations,''
  JETP Lett.\  {\bf 42}, 152 (1985)
  [Pisma Zh.\ Eksp.\ Teor.\ Fiz.\  {\bf 42}, 124 (1985)];
  %%CITATION = JTPLA,42,152;%%
  %382 citations counted in INSPIRE as of 25 Dec 2013
  %\cite{Sasaki:1995aw}
%\bibitem{Sasaki:1995aw} 
  M.~Sasaki and E.~D.~Stewart,
  %``A General analytic formula for the spectral index of the density perturbations produced during inflation,''
  Prog.\ Theor.\ Phys.\  {\bf 95}, 71 (1996)
  [astro-ph/9507001];
  %%CITATION = ASTRO-PH/9507001;%%
  %548 citations counted in INSPIRE as of 25 Dec 2013
  %\cite{Sasaki:1998ug}
%\bibitem{Sasaki:1998ug} 
  M.~Sasaki and T.~Tanaka,
  %``Superhorizon scale dynamics of multiscalar inflation,''
  Prog.\ Theor.\ Phys.\  {\bf 99}, 763 (1998)
  [gr-qc/9801017];
  %%CITATION = GR-QC/9801017;%%
  %224 citations counted in INSPIRE as of 25 Dec 2013
  %\cite{Lyth:2004gb}
%\bibitem{Lyth:2004gb} 
  D.~H.~Lyth, K.~A.~Malik and M.~Sasaki,
  %``A General proof of the conservation of the curvature perturbation,''
  JCAP {\bf 0505}, 004 (2005)
  [astro-ph/0411220];
  %%CITATION = ASTRO-PH/0411220;%%
  %400 citations counted in INSPIRE as of 25 Dec 2013
  
  %\cite{Sasaki:2006kq}
\bibitem{Sasaki:2006kq} 
  M.~Sasaki, J.~Valiviita and D.~Wands,
  %``Non-Gaussianity of the primordial perturbation in the curvaton model,''
  Phys.\ Rev.\ D {\bf 74}, 103003 (2006)
  [astro-ph/0607627].
  %%CITATION = ASTRO-PH/0607627;%%
  
    %\cite{Kofman:1995fi}
\bibitem{Kofman:1995fi} 
  L.~Kofman, A.~D.~Linde and A.~A.~Starobinsky,
  %``Nonthermal phase transitions after inflation,''
  Phys.\ Rev.\ Lett.\  {\bf 76}, 1011 (1996)
  [hep-th/9510119].
  %%CITATION = HEP-TH/9510119;%%
  %243 citations counted in INSPIRE as of 18 Aug 2015
  
  %\cite{Tkachev:1995md}
\bibitem{Tkachev:1995md} 
  I.~I.~Tkachev,
  %``Phase transitions at preheating,''
  Phys.\ Lett.\ B {\bf 376}, 35 (1996)
  [hep-th/9510146].
  %%CITATION = HEP-TH/9510146;%%
  %103 citations counted in INSPIRE as of 18 Aug 2015
  
       %\cite{Kawasaki:2013ae}
\bibitem{Kawasaki:2013ae} 
  For a review, see M.~Kawasaki and K.~Nakayama,
  %``Axions: Theory and Cosmological Role,''
  Ann.\ Rev.\ Nucl.\ Part.\ Sci.\  {\bf 63}, 69 (2013)
  [arXiv:1301.1123 [hep-ph]].
  %%CITATION = ARXIV:1301.1123;%%
  
    %\cite{Kasuya:1996ns}
\bibitem{Kasuya:1996ns} 
  S.~Kasuya, M.~Kawasaki and T.~Yanagida,
  %``Cosmological axion problem in chaotic inflationary universe,''
  Phys.\ Lett.\ B {\bf 409}, 94 (1997)
  [hep-ph/9608405];
  %%CITATION = HEP-PH/9608405;%%
  %\cite{Kasuya:1997td}
%\bibitem{Kasuya:1997td} 
  %S.~Kasuya, M.~Kawasaki and T.~Yanagida,
  %``Domain wall problem of axion and isocurvature fluctuations in chaotic inflation models,''
  Phys.\ Lett.\ B {\bf 415}, 117 (1997)
  [hep-ph/9709202];
  %%CITATION = HEP-PH/9709202;%%  
  %\cite{Kawasaki:2013iha}
%\bibitem{Kawasaki:2013iha} 
  M.~Kawasaki, T.~T.~Yanagida and K.~Yoshino,
  %``Domain wall and isocurvature perturbation problems in axion models,''
  JCAP {\bf 1311}, 030 (2013)
  [arXiv:1305.5338 [hep-ph]].
  %%CITATION = ARXIV:1305.5338;%%
  
   %\cite{Moroi:2013tea}
\bibitem{Moroi:2013tea} 
  T.~Moroi, K.~Mukaida, K.~Nakayama and M.~Takimoto,
  %``Scalar Trapping and Saxion Cosmology,''
  JHEP {\bf 1306}, 040 (2013)
  [arXiv:1304.6597 [hep-ph]];
  %%CITATION = ARXIV:1304.6597;%%
   %\cite{Moroi:2014mqa}
%\bibitem{Moroi:2014mqa} 
  %T.~Moroi, K.~Mukaida, K.~Nakayama and M.~Takimoto,
  %``Axion Models with High Scale Inflation,''
  JHEP {\bf 1411}, 151 (2014)
  [arXiv:1407.7465 [hep-ph]].
  %%CITATION = ARXIV:1407.7465;%%
  
  %\cite{Gherghetta:1995dv}
\bibitem{Gherghetta:1995dv} 
  T.~Gherghetta, C.~F.~Kolda and S.~P.~Martin,
  %``Flat directions in the scalar potential of the supersymmetric standard model,''
  Nucl.\ Phys.\ B {\bf 468}, 37 (1996)
  [hep-ph/9510370].
  %%CITATION = HEP-PH/9510370;%%
  
      %\cite{Murayama:1992dj}
\bibitem{Murayama:1992dj} 
  H.~Murayama, H.~Suzuki and T.~Yanagida,
  %``Radiative breaking of Peccei-Quinn symmetry at the intermediate mass scale,''
  Phys.\ Lett.\ B {\bf 291}, 418 (1992).
  %%CITATION = PHLTA,B291,418;%%
  
    %\cite{Kim:1986ax}
\bibitem{Kim:1986ax} 
  For a review, see J.~E.~Kim,
  %``Light Pseudoscalars, Particle Physics and Cosmology,''
  Phys.\ Rept.\  {\bf 150}, 1 (1987);
  %%CITATION = PRPLC,150,1;%%
  %\cite{Kim:2008hd}
%\bibitem{Kim:2008hd} 
  J.~E.~Kim and G.~Carosi,
  %``Axions and the Strong CP Problem,''
  Rev.\ Mod.\ Phys.\  {\bf 82}, 557 (2010)
  [arXiv:0807.3125 [hep-ph]].
  %%CITATION = ARXIV:0807.3125;%%
  
   %\cite{Linde:1990yj}
\bibitem{Linde:1990yj} 
  A.~D.~Linde and D.~H.~Lyth,
  %``Axionic domain wall production during inflation,''
  Phys.\ Lett.\ B {\bf 246}, 353 (1990);
  %%CITATION = PHLTA,B246,353;%%
  %\cite{Linde:1991km}
%\bibitem{Linde:1991km} 
  A.~D.~Linde,
  %``Axions in inflationary cosmology,''
  Phys.\ Lett.\ B {\bf 259}, 38 (1991).
  %%CITATION = PHLTA,B259,38;%%
   
%\cite{Kawasaki:2008mc}
\bibitem{Kawasaki:2008mc} 
  M.~Kawasaki, K.~Nakayama and F.~Takahashi,
  %``Hilltop Non-Gaussianity,''
  JCAP {\bf 0901}, 026 (2009)
  [arXiv:0810.1585 [hep-ph]].
  %%CITATION = ARXIV:0810.1585;%%
  
  %\cite{Kawasaki:2011pd}
\bibitem{Kawasaki:2011pd} 
  M.~Kawasaki, T.~Kobayashi and F.~Takahashi,
  %``Non-Gaussianity from Curvatons Revisited,''
  Phys.\ Rev.\ D {\bf 84}, 123506 (2011)
  [arXiv:1107.6011 [astro-ph.CO]].
  %%CITATION = ARXIV:1107.6011;%%

  
  
%%%%%%%%%%%%%%%%%%%%%%%%%%%%%%%%%%%%%%%%%%%%%%%%%%
\end{thebibliography}
\end{document}